\documentclass[twocolumn,aps,pra,showpacs,superscriptaddress,nofootinbib]{revtex4-1}

\usepackage[caption=false]{subfig}
\usepackage{amsfonts}
\usepackage{filecontents}
\usepackage{graphicx}
\usepackage{lipsum}
\usepackage{mathtools}
\usepackage{microtype}
\usepackage{paralist}
\usepackage{pgfplots}
\usepackage{physics}
\usepackage{siunitx}

\usepackage{hyperref}
\usepackage{cleveref}

\pgfplotsset{compat=1.4}

\usepgfplotslibrary{external}
\usepgfplotslibrary{colorbrewer}
\usetikzlibrary{pgfplots.groupplots}

\newcommand{\qd}{quantum dot}
\newcommand{\qds}{quantum dots}

\newcommand{\outerprod}[2]{#1 \! \otimes \! #2}

\begin{document}

\title{Collective Rabi dynamics of electromagnetically-coupled quantum dot ensembles}

\author{Connor Glosser}
\email[Corresponding author; ]{glosser1@msu.edu}
\affiliation{Department of Physics \& Astronomy, Michigan State University\\567 Wilson Road, East Lansing MI 48824 USA}
\affiliation{Department of Electrical \& Computer Engineering, Michigan State University\\428 South Shaw Lane, East Lansing MI 48824, USA}

\author{B.\ Shanker}
\affiliation{Department of Electrical \& Computer Engineering, Michigan State University\\428 South Shaw Lane, East Lansing MI 48824, USA}

\author{Carlo Piermarocchi}
\email[ ]{piermaro@msu.edu}
\affiliation{Department of Physics \& Astronomy, Michigan State University\\567 Wilson Road, East Lansing MI 48824 USA}

\date{\today}
\pacs{1.1}

\begin{abstract}
  Rabi oscillations typify the inherent nonlinearity of optical excitations in \qds{}.
Using an integral kernel formulation to solve the 3D Maxwell-Bloch equations in ensembles of up to $10^4$ quantum dots, we observe features in Rabi oscillations due to the interplay of nonlinearity, non-equilibrium excitation, and electromagnetic coupling between the dots.
This approach allows us to observe the dynamics of each dot in the ensemble without resorting to spatial averages.
Our simulations predict synchronized multiplets of dots that exchange energy, dots that dynamically couple to screen the effect of incident external radiation, localization of the polarization due to randomness and interactions, as well as wavelength-scale regions of enhanced and suppressed polarization.

\end{abstract}

\maketitle

\section{\label{section:introduction}Introduction}

Semiconductor structures containing a large number of \qds{} offer ideal environments for exploring collective effects induced by light-matter interactions.
Often, these structures exhibit new phenomena due to geometrical randomness and nonlinearities in the underlying system dynamics.
Additionally, optical excitations (excitons) undergo characteristic Rabi oscillations~\cite{Stievater2001,Kamada2001,Htoon2002} in \qds{} analogous to those observed in atomic systems; as \qds{} have stronger dipolar transitions than atoms, these light-induced oscillations generate secondary fields that couple the system more strongly than equivalent atomic species.
We can therefore expect---at least in some regions of the sample---these local secondary fields will produce modified collective behavior in the exciton dynamics.
Phenomena induced by these secondary fields have received considerable theoretical/computational~\cite{Slepyan2002,Slepyan2004} and experimental~\cite{Asakura2013} attention as they may provide new insight on the coherent dynamics of excitons in \qd{} systems.

In the realm of theoretical/computational investigation, researchers in atomic and solid-state optics have developed numerous variations of the Maxwell-Bloch equations~\cite{Gross1982} to describe features such as ringing in pulse propagation~\cite{Burnham1969,MacGillivray1976} or emission fluctuations~\cite{Haake1979}.
Early solution strategies for these equations fell to continuum models~\cite{Rehler1971,MacGillivray1976} that recover effects arising from far-field interactions, but cannot adequately describe near-field regimes.
More recently, mesh-based PDE solvers~\cite{Vanneste2001,Fratalocchi2008,Bachelard2015} added a large degree of fidelity to these models, though the finite size of the mesh means they still have trouble resolving short-range effects without unduly increasing the computational cost.
Additionally, the nature of these meshes make them prohibitively expensive to extend into higher dimensional geometries for optically-large systems.
In this work we develop a computational framework to discover signatures of collective effects in strongly-driven \qds{} within a microscopic formalism.
By constructing the Maxwell-Bloch equations with an integral kernel to describe radiation, we recover near- and far-electric fields with full fidelity across the simulation while allowing for dynamics at the level of individual \qds{}.
Our methodology---based on successful models of other electromagnetic~\cite{Shanker2000,Pray2012,Pray2014} and acoustic~\cite{Ergin1999a,Ergin1999b,Glosser2016} systems---accommodates $10^4$ particles distributed over optically-large regions in three dimensions.

As we explicitly track the evolution of each \qd{} in the system, we will numerically demonstrate that the collective Rabi oscillation can induce significant coupling in sufficiently close \qds{}.
This laser-induced inter-dot coupling manifests itself in different forms:
\begin{inparaenum}[(i)]
  \item The polarization generated in isolated \qd{} pairs dynamically suppresses the Rabi rotation.
  We interpret this as the consequence of a time-dependent energy shift that brings the pair temporarily out of resonance with the external driving field.
  \item In addition to this screening, we observe oscillations in the free-induction decay for larger multiplets of \qds{}.
  \item Optical pulses of integer $\pi$ area, for which we expect no polarization in uncoupled systems following the pulse, produce patterns of residual localized polarization that remain in the system.
  \item The long-range interactions in optically-large systems produce wavelength-scale regions of enhanced and suppressed polarization.
\end{inparaenum}
These effects could, for instance, help identify multiplets of dots that dynamically couple during Rabi oscillations, or help understand nonlinear pulse propagation effects in these media.

We structure the remainder of this paper as follows: \cref{section:problem statement} motivates the physical model of an ensemble of two-level systems that interact through a classical electric field.
\Cref{section:computational approach} presents the details of our methodology in the context of a global rotating-wave approximation and we offer an implementation of this algorithm at~\cite{quest_release}.
\Cref{section:results} contains the results of our investigation where we observe polarization features not present in noninteracting systems at both sub- and super-wavelength scales.
Finally, \cref{section:conclusion} contains concluding remarks where we hypothesize on the mechanisms underpinning the observed polarization features as well as comment on our future work in this area.

\section{\label{section:problem statement}Problem Statement}
Consider the evolution of a set of \qds{} in response to a time-varying electric field.
If we concern ourselves only with electric dipole transitions in a resonant (or nearly-resonant) system, we may write the time-dependence of a given \qd's density matrix, $\hat{\rho}(t)$, as
\begin{equation}
  \dv{\hat{\rho}}{t} = \frac{-i}{\hbar}\commutator{\hat{\mathcal{H}}(t)}{\hat{\rho}} - \hat{\mathcal{D}}\qty[\hat{\rho}].
  \label{eq:liouville}
\end{equation}
Here, $\hat{\mathcal{H}}(t)$ represents a local Hamiltonian that governs the internal two-level structure of the \qd{}, as well as its interaction with an external electromagnetic field, and $\hat{\mathcal{D}}$ provides dissipation terms that account for emission effects phenomenologically.
Formally,
\begin{subequations}
  \begin{align}
    \hat{\mathcal{H}}(t) &\equiv \mqty(0 & \hbar \chi(t) \\ \hbar \chi^*(t) & \hbar \omega_0) \label{eq:hamiltonian}\\
    \hat{\mathcal{D}}\qty[\hat{\rho}] &\equiv \mqty( \qty(\rho_{00} - 1)/{T_1} & \rho_{01}/{T_2} \\ \rho_{10}/{T_2} & \rho_{11}/T_1 ) \label{eq:dissipator}
  \end{align}
\end{subequations}
where $\chi(t) \equiv \vb{d} \cdot \hat{\vb{E}}(\vb{r}, t)/\hbar$, $\vb{d} \equiv \matrixel{1}{e \hat{\vb{r}}}{0}$, and $\ket{0}$ \& $\ket{1}$ represent the highest valence and lowest conduction states of the \qd{} under consideration.
Finally, the $T_1$ and $T_2$ constants characterize average emission and relaxation times.

To account for the interactions between \qds{}, we turn to a semiclassical description of the system under the assumption of coherent fields and negligible quantum  statistics effects.
Such an approximation preserves the discrete two-level energy structure of \emph{individual} \qds{} though electromagnetic quantities behave like their classical analogues.
We define the total electric field at any point as $\vb{E}(\vb{r}, t) = \vb{E}_L(\vb{r}, t) + \mathfrak{F}\{ \vb{P}(\vb{r}, t) \}$
where $\vb{E}_L(\vb{r}, t)$ describes an incident laser field, $\vb{P}(\vb{r}, t)$ a polarization distribution arising from  the off-diagonal elements (coherences) of $\hat{\rho}$, and
\begin{equation}
  \begin{gathered}
    \mathfrak{F}\{\vb{P}(\vb{r}, t)\} \equiv
      \frac{-1}{4\pi \epsilon} \int
      \left(\tensor{\mathrm{I}} -  \outerprod{\bar{\vb{r}}}{\bar{\vb{r}}} \right) \cdot \frac{\partial_t^2 \vb{P}(\vb{r}', t_R)}{c^2 \abs{\vb{r}-\vb{r}'}} + \\
      \left(\tensor{\mathrm{I}} - 3\outerprod{\bar{\vb{r}}}{\bar{\vb{r}}} \right) \cdot \qty(
        \frac{\partial_t   \vb{P}(\vb{r}', t_R)}{c \abs{\vb{r}-\vb{r}'}^2} +
        \frac{             \vb{P}(\vb{r}', t_R)}{  \abs{\vb{r}-\vb{r}'}^3}
      ) \dd[3]{\vb{r}}'
  \end{gathered}
  \label{eq:integral operator}
\end{equation}
(see \S 72 of \cite{Landau2013}).
Here, $\tensor{\mathrm{I}}$ denotes the identity dyad, $\bar{\vb{r}} \equiv \qty(\vb{r} - \vb{r}')/\abs{\vb{r} - \vb{r}'}$, $\otimes$ represents the tensor product (i.e.\ $\qty(\outerprod{\vb{a}}{\vb{b}})_{ij} = a_i b_j$), $t_R \equiv t - \abs{\vb{r} - \vb{r}'}/c$, and $\epsilon$ gives the dielectric constant of the inter-dot medium.
Thus, in a system composed of multiple \qds{}, \cref{eq:integral operator} couples the evolution of each \qd{} by way of the off-diagonal matrix elements appearing in \cref{eq:hamiltonian}.
Note that this approach does not require an instantaneous dipole-dipole Coulomb term between (charge-neutral) \qds{}; the interactions between structures occur only via the electric field which propagates through space with finite velocity.
(See {A}$_{\textsc{iv}}$  and {C}$_{\textsc{iv}}$ of~\cite{Cohen1989} for in-depth discussions of this point.)

In the systems under consideration here, $\omega_0$ lies in the optical frequency band ($\sim \SI{1500}{\milli\eV\per\hbar}$).
As such, na\"ively integrating \cref{eq:liouville} to resolve the Rabi dynamics that occur on the order of \SI{1}{\pico\second} becomes computationally infeasible.
By introducing $\tilde{\rho} = \hat{U} \hat{\rho} \hat{U}^\dagger$ where $\hat{U} = \mathrm{diag}(1, e^{i \omega_L t})$, we may instead write \cref{eq:liouville} as
\begin{equation}
  \dv{\tilde{\rho}}{t} = \frac{-i}{\hbar} \commutator{\hat{U} \hat{\mathcal{H}} \hat{U}^\dagger - i \hbar \hat{V}}{\tilde{\rho}} - \hat{\mathcal{D}}\qty[\tilde{\rho}], \quad \hat{V} \equiv \hat{U} \dv{\hat{U}^\dagger}{t}
  \label{eq:rotating liouville}
\end{equation}
which will contain only terms proportional to $e^{i (\omega_0 \pm \omega_L) t}$ if $\vb{E}(t) \sim \tilde{\vb{E}}(t)\cos(\omega_L t)$.
Consequently, we ignore the high-frequency quantities under the assumption that such terms will integrate to zero in solving \cref{eq:rotating liouville} over appreciable timescales~\cite{Allen1975}. As the system no longer contains any optical frequencies, one can then construct efficient numerical strategies for solving \cref{eq:rotating liouville}.

Due to the quantum mechanical transitions at play in producing secondary radiation, we may assume similarly monochromatic radiated fields.
As such, a similar transformation applies to the source distribution in \cref{eq:integral operator}.
Writing $\vb{P}(\vb{r}, t) = \tilde{\vb{P}}(\vb{r}, t)e^{i \omega_L t}$ and similarly ignoring high-frequency terms, the radiated field envelope becomes
\begin{widetext}
\begin{equation}
  \begin{gathered}
    \tilde{\mathfrak{F}}\{ \tilde{\vb{P}}(\vb{r}, t) \} \equiv \frac{-1}{4\pi \varepsilon} \int
    \qty(\tensor{\mathrm{I}} -  \outerprod{\bar{\vb{r}}}{\bar{\vb{r}}} ) \cdot \frac{\qty(\partial_t^2 \tilde{\vb{P}}(\vb{r}', t_R) + 2 i \omega_L \partial_t \tilde{\vb{P}}(\vb{r}', t_R) - \omega_L^2 \tilde{\vb{P}}(\vb{r}', t_R)) e^{-i \omega_L \abs{\vb{r} - \vb{r}'}/c}}{c^2 \abs{\vb{r}-\vb{r}'}} + \\
    \qty(\tensor{\mathrm{I}} - 3\outerprod{\bar{\vb{r}}}{\bar{\vb{r}}} ) \cdot \frac{\qty(\partial_t \tilde{\vb{P}}(\vb{r}', t_R) + i \omega_L \tilde{\vb{P}}(\vb{r}', t_R))e^{-i \omega_L \abs{\vb{r} - \vb{r}'}/c}}{c \abs{\vb{r}-\vb{r}'}^2} +
    \qty(\tensor{\mathrm{I}} - 3\outerprod{\bar{\vb{r}}}{\bar{\vb{r}}} ) \cdot \frac{                \tilde{\vb{P}}(\vb{r}', t_R) e^{-i \omega_L \abs{\vb{r} - \vb{r}'}/c}}{\abs{\vb{r}-\vb{r}'}^3}
    \, \dd[3]{\vb{r'}}.
  \end{gathered}
  \label{eq:radiated envelope}
\end{equation}
\end{widetext}
Critically, \cref{eq:radiated envelope} maintains the high-frequency phase relationship between sources oscillating at $\omega_L$ via the factors of $e^{-i \omega_L \abs{\vb{r} - \vb{r}'}/c}$ that appear.

\section{\label{section:computational approach}Computational Approach}

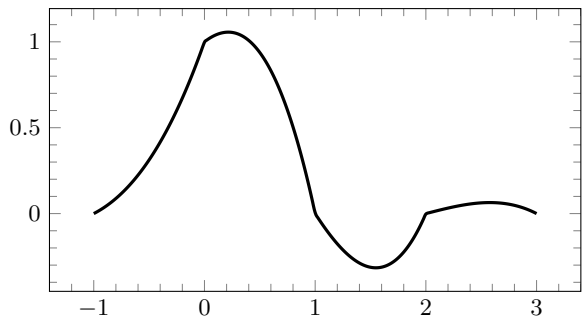
\begin{figure}
  \begin{tikzpicture}[
  declare function = {
    basis(\x) = 
      and(-1 <= \x, \x < 0)*(1+x)*(2+x)*(3+x)/6 +
      and(0  <= \x, \x < 1)*(1-x)*(1+x)*(2+x)/2 +
      and(1  <= \x, \x < 2)*(1-x)*(2-x)*(1+x)/2 + 
      and(2  <= \x, \x < 3)*(1-x)*(2-x)*(3-x)/6 + 
      or(\x <-1, \x > 3)*0;
  }
  ]

  \begin{axis}[width=\columnwidth, height=0.61803398875\columnwidth,
      minor x tick num={4},
      minor y tick num={4},
  ]
  
  \addplot[very thick, domain=-1:3, smooth, samples=256] {basis(x)};
  \end{axis}
\end{tikzpicture}
  \caption{\label{fig:interpolation basis} Nonorthogonal and $C^0$-continuous temporal basis function $T(t)$ constructed from intervals of third-order Lagrange polynomials.}
\end{figure}
To solve \cref{eq:rotating liouville,eq:radiated envelope} for each of $N_s$ \qds{} at $N_t$ equally-spaced timesteps, we begin with a suitable representation of $\tilde{\vb{P}}(\vb{r}, t)$ in terms of spatial and temporal basis functions, i.e.~
\begin{equation}
  \tilde{\vb{P}}(\vb{r}, t) = \sum_{\ell=0}^{N_s - 1} \sum_{m = 0}^{N_t - 1} \mathcal{A}_{\ell}^{(m)} \vb{S}_\ell(\vb{r}) T(t - m \, \Delta t).
  \label{eq:basis function representation}
\end{equation}
As the wavelength of any radiation in the system far exceeds the dimensions of the \qds{} under consideration, we take $\vb{S}_\ell(\vb{r}) \equiv \vb{d}_\ell \delta(\vb{r} - \vb{r}_\ell)$ where $\vb{d}_\ell$ and $\vb{r}_\ell$ denote the dipole moment and position of dot $\ell$.
Furthermore, we require the $T(t)$ to have finite support as well as causal and interpolatory properties so as to recover $\tilde{\vb{P}}$, $\partial_t \tilde{\vb{P}}$, and $\partial^2_t \tilde{\vb{P}}$ at every timestep.
Accordingly, we have elected to use
\begin{equation}
  T(t) = \sum_{j = 0}^p \lambda_j(t)
  \label{eq:basis sum}
\end{equation}
where
\begin{equation}
  \lambda_j(t) =
  \begin{cases}
    \frac{(1 - \tau)_j}{j!} \frac{(1 + \tau)_{p-j}}{(p-j)!} & j-1 \le \tau < j \\
    0 & \text{otherwise,}
  \end{cases}
  \label{eq:basis piece}
\end{equation}
$(a)_k \equiv \Gamma(a + k)/\Gamma(a)$ denotes the Pochhammer rising factorial, and $\tau \equiv t/\Delta t$.
Such a $T(t)$ consists of shifted, backwards-looking Lagrange polynomials of order $p$ (we require $p \ge 3$ to recover a twice-differentiable function), forming a temporal basis set with functions similar to the one shown in \cref{fig:interpolation basis}.
These functions  reliably interpolate smooth functions with controllable error and have a long history of use in studies of radiative systems~\cite{Manara1997,Bluck1997}.

Combining \cref{eq:radiated envelope,eq:basis function representation} and projecting the resulting field onto the spatiotemporal basis functions produces a marching-on-in-time system of the form
\begin{equation}
  \mathcal{L}^{(m)} + \sum_{k = 0}^{m} \mathcal{Z}^{(k)} \mathcal{A}^{(m - k)} = \mathcal{F}^{(m)}.
  \label{eq:zmatrix}
\end{equation}
In this (block $N_s \times N_s$) matrix equation
\begin{subequations}
  \begin{align}
    \mathcal{L}^{(m)}_{\ell} &= \big\langle \mathbf{S}_\ell(\vb{r}), \tilde{\vb{E}}_L(\vb{r}, m \, \Delta t) \big\rangle \\
    \mathcal{Z}^{(k)}_{\ell \ell'} &= \big\langle \mathbf{S}_\ell(\vb{r}), \tilde{\mathfrak{F}}\{ \mathbf{S}_{\ell'}(\mathbf{r}) T(k \, \Delta t) \} \big\rangle \\
    \mathcal{F}^{(m)}_\ell &= \big\langle \mathbf{S}_\ell(\vb{r}), \tilde{\mathbf{E}}(\vb{r}, m \, \Delta t) \big\rangle
  \end{align}
  \label{eq:projections}
\end{subequations}
where $\langle \cdot, \cdot \rangle$ denotes the scalar product of two functions.
As we have elected to use a uniform $\Delta t$, the summation in \cref{eq:zmatrix} represents a discrete convolution that will produce $\mathcal{F}^{(m)}$ given only $\mathcal{A}^{(m' \le m)}$.
To link $\mathcal{A}^{(m)}$ with the density matrix in \cref{eq:rotating liouville}, we take
\begin{equation}
  \mathcal{A}^{(m)}_\ell \equiv \tilde{\rho}_{\ell, 01}(t = m \, \Delta t)
  \label{eq:polarization definition}
\end{equation}
as the off-diagonal matrix elements (coherences) of $\tilde{\rho}_{\ell}$ directly characterize the dipole radiating at $\vb{r}_\ell$ under the rotating-wave approximation.
Consequently, determining $\mathcal{A}^{(m + 1)}$ amounts to integrating \cref{eq:rotating liouville} from $t_i = m \, \Delta t$ to $t_f = \qty(m + 1) \, \Delta t$ for every \qd{} in the system.
For this, we make use of the predictor/corrector scheme detailed in~\cite{Glaser2009}.
Defining $t_m \equiv m \, \Delta t$ and approximating $\tilde{\rho}(t)$ as a weighted sum of complex exponentials, the predictor/corrector scheme proceeds with an extrapolation predictor step,
\begin{equation}
  \tilde{\rho}_\ell(t_{m + 1}) \leftarrow \sum_{w = 0}^{W-1} \mathcal{P}_w^{\qty(0)} \tilde{\rho}_\ell(t_{m-w}) + \mathcal{P}_w^{\qty(1)} \, \partial_t \tilde{\rho}_\ell(t_{m-w}),
  \label{eq:predictor}
\end{equation}
and iterated corrector steps,
\begin{equation}
    \tilde{\rho}_\ell(t_{m + 1}) \leftarrow \sum_{w = -1}^{W - 1} \mathcal{C}_w^{\qty(0)} \tilde{\rho}_\ell(t_{m-w}) + \mathcal{C}_w^{\qty(1)} \, \partial_t \tilde{\rho}_\ell(t_{m-w})
  \label{eq:corrector}
\end{equation}
($\mathcal{C}_{-1}^{(0)} \equiv 0$ by construction) to advance the system by $\Delta t$.
Such an integrator has significantly better convergence properties than Runge-Kutta integrators for equations of the type seen in \cref{eq:rotating liouville}.

To summarize, one timestep of our solution strategy proceeds as follows:
\begin{enumerate}
  \item At timestep $m$, use \cref{eq:predictor,eq:polarization definition} to predict $\mathcal{A}^{(m + 1)}$.
    This prediction depends only on the known history of the system and does not require the calculation of any electromagnetic interactions.
  \item Use \cref{eq:zmatrix} to calculate $\mathcal{F}^{(m + 1)}$.
    Having extrapolated $\mathcal{A}^{(m + 1)}$ in step 1, $\mathcal{F}^{(m + 1)}$ will contain information from \qds{} within $c \, \Delta t$ of each other.
  \item Produce $\partial_t \tilde{\rho}(t_{m + 1})$ (and thus $\partial_t \mathcal{A}^{(m + 1)}$) by evaluating \cref{eq:rotating liouville} with the $\mathcal{F}^{(m+1)}$ found in step 2.
  \item Correct $\mathcal{A}^{(m + 1)}$ with \cref{eq:corrector} and $\partial_t \mathcal{A}^{(m + 1)}$ found in step 3.
    Repeat steps 2 through 4 until $\mathcal{A}^{(m + 1)}$ has sufficiently converged, then set $m \leftarrow m + 1$.
\end{enumerate}

\section{\label{section:results}Numerical Results}
\begin{table}
  \begin{ruledtabular}
    \begin{tabular}{lll}
      Quantity                 & Symbol            & Value                        \\ \hline
      Speed of light           & $c$               & \SI{300}{\micro\meter \per \pico\second} \\
      Transition frequency     & $\omega_0$        & $\SI{1500}{\milli\eV}/\hbar$ \\
      Transition dipole moment & $\vb{d}$          & \SI{10}{\elementarycharge\bohr} (uniform) \\
      Decoherence times        & $T_{1}, T_{2}$    & \SIlist{10;20}{\pico\second} \\
      Laser frequency          & $\omega_L$        & $\SI{1500}{\milli\eV}/\hbar$ \\
      Laser wavevector         & $\vb{k}_L$        & $\omega_L/c$ ($\vb{k}_L \cdot \vb{d} = 0$) \\
      Pulse width              & $\sigma/\omega_L$ & \SI{1}{\pico\second} \\
      Pulse area               &  --               & $\pi$ \\
      %Number density          & $n$               & $< \SI{1.6e4}{\micro\meter\tothe{-3}}$ \\
      %\hline
      %Speed of light           & $c$            & \SI{299.792458}{\micro\meter\per\pico\second} \\
      %Reduced Planck constant  & $\hbar$        & \SI{0.65821193}{\milli\eV \pico\second} \\
      %Vacuum permeability      & $\mu_0$        & \SI{2.013e-4}{\milli\eV \pico\second\squared \per \elementarycharge \per \micro\meter}
    \end{tabular}
  \end{ruledtabular}
  \caption{\label{table:parameters}Simulation parameters (unless otherwise stated); \si{\elementarycharge} and \si{\bohr} denote the elementary charge and Bohr radius.
    The decoherence times here, while shorter than those typical of optical resonance experiments, afford a shorter computational time but preserve dynamical emission phenomena.
  }
\end{table}

Here we detail the results of investigations into coupled \qd{} behavior with the model presented thus far.
Our algorithm reliably handles tens of thousands of \qds{} and can simulate ten picoseconds of system dynamics in two days on a single processor.
We perform simulations of systems of \qds{} randomly distributed throughout a simulation volume experiencing a laser pulse of the form
\begin{equation}
  \tilde{\mathbf{E}}_L(\vb{r}, t) = \tilde{\mathbf{E}}_0 e^{-(\vb{k}_L \cdot \vb{r} - \omega_L t)^2/(2\sigma^2)}.
  \label{eq:pulse envelope}
\end{equation}
\Cref{table:parameters} provides the physical system parameters unless otherwise stated.

\subsection{Stability \& adjacency effects}

\begin{figure}
  \usepgfplotslibrary{colorbrewer}
\usepgfplotslibrary{fillbetween}
\usetikzlibrary{patterns}
\usetikzlibrary{pgfplots.groupplots}
\usetikzlibrary{spy}

\begin{tikzpicture}[
    spy using outlines={circle, magnification=12, connect spies},
  ]
  \begin{groupplot}[
      group style={
          group name=my plots,
          group size=1 by 2,
          xlabels at=edge bottom,
          xticklabels at=edge bottom,
          vertical sep=2pt,
      },
      width=\columnwidth,
      height=0.618034\columnwidth,
      xlabel={Time (\si{\pico\second})},
      xmin = 0, xmax = 30,
      ymin = 0, ymax = 1,
      minor x tick num={4},
      minor y tick num={4},
      enlargelimits = 0.10,
      legend pos=south east,
      cycle list/Set1-3
  ]
  \nextgroupplot[ylabel style={align=center}, ylabel={Population\\(pair only)}]
    \addplot+[smooth, thick, no marks] table[x = time, y = p1] {pair_density_stats.dat};
    \addlegendentry{$\rho_{00}^{\qty(1)}(t)$}

    \addplot+[smooth, thick, no marks] table[x = time, y = p2] {pair_density_stats.dat};
    \addlegendentry{$\rho_{00}^{\qty(2)}(t)$}

    \addplot[smooth, thick, dashed, no marks] table[x = time, y = lone] {pair_density_stats.dat};
    \addlegendentry{$\tilde{\vb{E}}_L$-only}

    \node[] at (axis cs: 0,0) {(a)};

  \nextgroupplot[ylabel style={align=center}, ylabel={Population\\(neighborhood)}]
    \addplot+[smooth, thick, no marks] table[x = time, y = p1] {high_density_stats.dat};
    \addlegendentry{$\rho_{00}^{\qty(1)}(t)$}

    \addplot+[smooth, thick, no marks] table[x = time, y = p2] {high_density_stats.dat};
    \addlegendentry{$\rho_{00}^{\qty(2)}(t)$}

    \foreach \n in {5, ..., 46}
    {
      \addplot[smooth, thick, opacity = 0.2] table[x = time, y index={\n}] {high_density_stats.dat};
    }
    \addlegendentry{Neighborhood}

    \node[] at (axis cs: 0,0) {(b)};

  \end{groupplot}

\end{tikzpicture}
  \caption{\label{fig:density stats}Population dynamics of adjacent \qds{} (top, shown with the trajectory of a \qd{} driven exclusively by the external laser in black) and a 1024-particle neighborhood (dots (1) and (2) maintain their separation and orientation between simulations).
    The interaction between particles gives a greatly diminished response to the external pulse through a dynamical detuning of the two-dot system.
    The majority of the neighboring particles in the 1024-dot system follow trajectories nearly identical to that prescribed by the laser (omitted for clarity); many-dot effects, however, produce significant oscillatory modes in the evolution of selected \qds{} (shown in gray).
    Note that $\rho_{00}^{\qty(1)}$ and $\rho_{00}^{\qty(2)}$ appear to have \emph{two} coherent modes in the presence of multiple dots: a high-frequency oscillation between the pair, as well as a low-frequency oscillation of the group about a decaying envelope.
}
\end{figure}
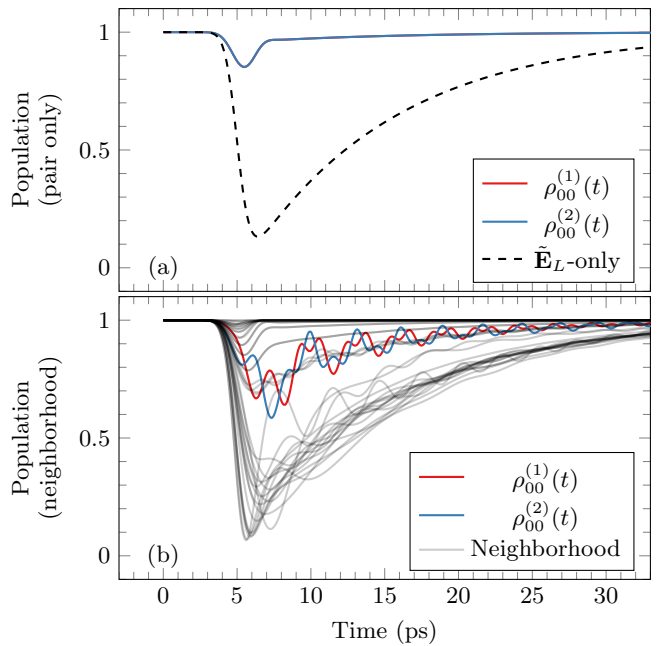

\Cref{fig:density stats} details $\rho_{00}(t)$ for two-, and 1024-particle simulations shown against a solution of \cref{eq:rotating liouville} for a \qd{} evolving according to $\tilde{\vb{E}}_L(\vb{r}, t)$ alone.
The system in \cref{fig:density stats}(a) contains two \qds{} with a separation of \SI{6.3}{\nano\meter} perpendicular to their mutual $\vb{d}$ to ensure large contributions from the near-field term of \cref{eq:radiated envelope}.
The two \qds{} in this system follow the same trajectory and both excite far less than either would in response to the incident laser alone.
These signatures indicate the effect arises as a dynamical frqeuency shift that brings adjacent \qds{} out of resonance with the applied electric field.
We note that this suppression effect occurred to varying degrees for all near-field arrangements of two \qds{} that we investigated.
In \cref{fig:density stats}(b), the system contains the same two-dot arrangement as in (a), however we have added an additional 1024 \qds{} randomly distributed throughout a $\SI{6.4e-2}{\micro\meter\cubed}$ cube centered around the original pair.
Most of these additional \qd{} populations deviate little from those produced by the laser-only pulse due to the large separation between particles.
Nevertheless, as we have filled the cube randomly, some regions of the system contain localized clusters of \qds{} that produce the suppression effect detailed above (for two adjacent particles) or populations with higher frequency oscillations (in the case of clusters with three or more \qds{}).
Specifically, the two ``original'' \qds{} acquire an out-of-phase oscillation with respect to each other as well as a lower frequency in-phase oscillation of the pair about a decaying envelope.

\begin{figure}
  \centering
  \begin{tikzpicture}
  \pgfplotsset{
    colormap={spectral}{
      rgb255=(94, 79, 162)
      rgb255=(50, 136, 189)
      rgb255=(102, 194, 165)
      rgb255=(171, 221, 164)
      rgb255=(230, 245, 152)
      rgb255=(255, 255, 191)
      rgb255=(254, 224, 139)
      rgb255=(253, 174, 97)
      rgb255=(244, 109, 67)
      rgb255=(213, 62, 79)
      rgb255=(158, 1, 66)
    }
  }
  \begin{axis}[grid=major,
    xmin = -0.2, xmax = 0.2,
    ymin = -0.2, ymax = 0.2,
    zmin = -0.2, zmax = 0.2,
    x dir=reverse, %indicate left-handed coordinate system
    3d box,
    colormap/bluered,
    colorbar,
    colorbar style = {
      ymin = 0,
      ymax = 0.5,
      plot graphics/ymin=0,
      plot graphics/ymax=0.5,
      colormap name=spectral
    }
  ]
    \addplot3 graphics[points={% important
        (0.15, 0.15, -0.15) => (1836, 2072-581)
        (0.15, 0.15, 0.15)  => (948, 2072-840)
        (0.15, -0.15, 0.15) => (948, 2072-1899)
        (-0.10, 0.15, 0.15) => (357, 2072-571)
      }] {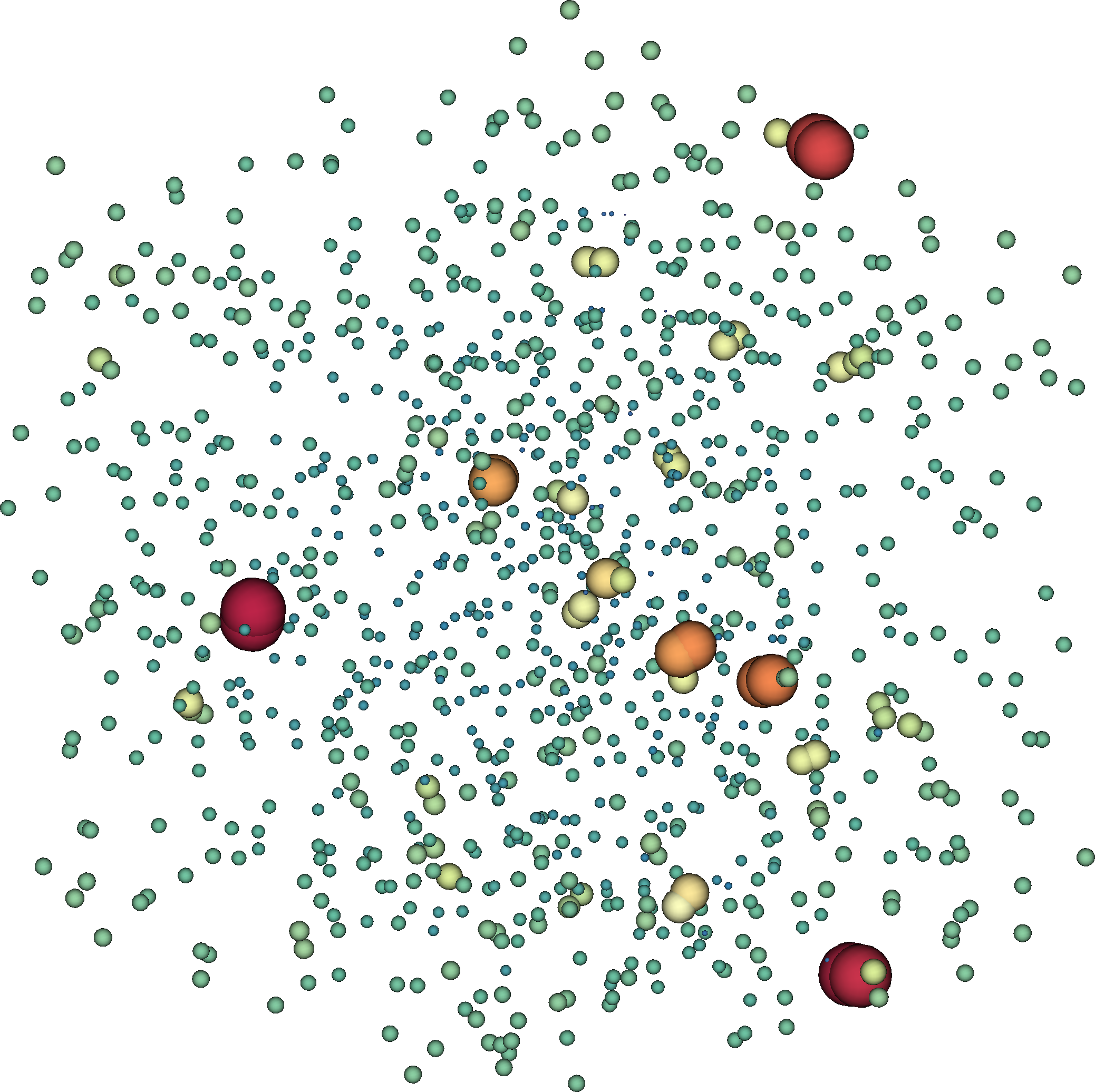};
  \end{axis}
\end{tikzpicture}
  \caption{\label{fig:nearfield box}Spatial distribution of $\abs{\tilde{\rho}_{01}}$ for 1024 \qds{} as an indicator of polarization.
    Recorded at $t = \SI{2}{\pico\second}$ relative to the peak of a \SI{1}{\pico\second}-wide $\pi$-pulse, the color and size of each sphere indicates the location of each \qd{} and its polarization.
    Following a $\pi$-pulse, a single \qd{} would have no remnant polarization; here, due to the near-field interactions between particles, clusters of \qds{} remain in highly-polarized states depending on their separation.
  }
\end{figure}

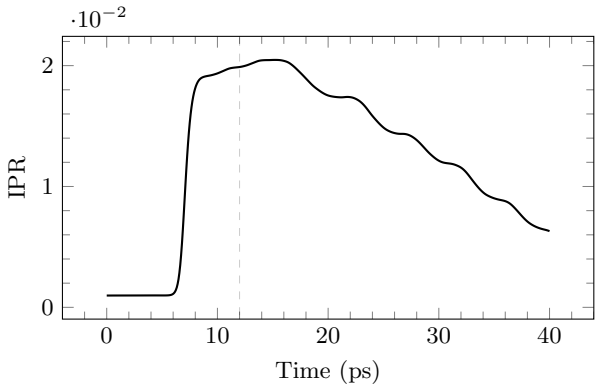
\begin{figure}
  \usetikzlibrary{pgfplots.groupplots}

\begin{tikzpicture}[trim axis left]
  \begin{axis}[
      width=\columnwidth,
      height=0.618034\columnwidth,
      xlabel={Time (\si{\pico\second})},
      ylabel={IPR},
      minor x tick num={4},
      minor y tick num={4},
      enlargelimits = 0.1
  ]
    \addplot[smooth, thick, no marks] table[x = {time}, y={ipr}] {ipr.dat};
    \draw[dashed, opacity=0.2] (axis cs:12,-1) -- (axis cs:12,1) {};
  \end{axis}
\end{tikzpicture}
  \caption{\label{fig:ipr}Inverse participation ratio (IPR) for the system shown in \cref{fig:nearfield box}. The dashed line indicates the sample time for \cref{fig:nearfield box}.}
\end{figure}

\Cref{fig:nearfield box} further illustrates the near-field coupling between \qds{}.
Here, 1024 \qds{} randomly fill an \SI{8e-3}{\micro\meter\cubed} cube and the same \SI{1}{\pico\second} Gaussian $\pi$-pulse illuminates the system.
Without any inter-dot interactions, such a pulse would perfectly transition every \qd{} from $\ket{0}$ to $\ket{1}$, leaving behind no polarization after the pulse has passed.
As the system in \cref{fig:nearfield box} has experienced most of the pulse, the majority of the \qds{} behave this way.
A number of particles remain polarized well after the pulse has passed through the system, however, due to their apparent proximity.

\subsection{Polarization enhancement}

Borrowing from standard measures of localization (in which one calculates integrals of $\vb{E}$ over the simulation volume), we have adapted the inverse participation ratio (IPR)  of the dot polarization as
\begin{equation}
  \text{IPR}(t) = \frac{\sum_\ell \abs{\tilde{\vb{p}}_\ell(t)}^4}{\qty(\sum_\ell \abs{\tilde{\vb{p}}_\ell(t)}^2)^2}
  \label{eq:ipr}
\end{equation}
to provide a quantitative description of these phenomena~\cite{Schwartz2007}.
\Cref{fig:ipr} shows this quantity for the system in \cref{fig:nearfield box}.
The maximum IPR occurs some time after the pulse has passed through the system, indicating strongly-coupled \qds{} retain their polarization longer than their neighbors.
Moreover, this dynamical localization effect features oscillations which suggests many-dot effects contribute to the dynamics within a narrow spectral region.

\begin{figure*}
  \centering
  \subfloat{
    \begin{tikzpicture}
      \begin{axis}[
          hide axis,
          scale only axis,
          height=0pt,
          width=0pt,
          colormap/viridis,
          colorbar horizontal,
          point meta min=0.32,
          point meta max=0.40,
          colorbar style={
              width=0.6180339887\textwidth,
              xtick={0.32, 0.33, 0.34, 0.35, 0.36, 0.37, 0.38, 0.39, 0.40},
              xticklabels={$\le 0.32$, $0.33$, $0.34$, $0.35$, $0.36$, $0.37$, $0.38$, $0.39$, $0.40 \le$}
          }]
          \addplot [draw=none] coordinates {(0,0)};
      \end{axis}
    \end{tikzpicture}
  } \\
  \subfloat{\includegraphics[width=\textwidth]{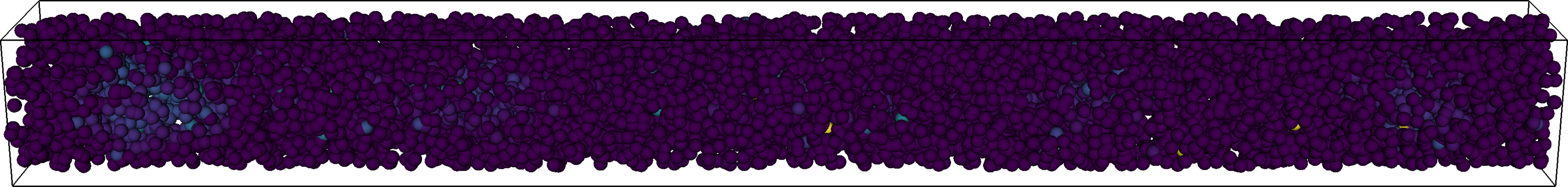}} \\
  \subfloat{\includegraphics[width=\textwidth]{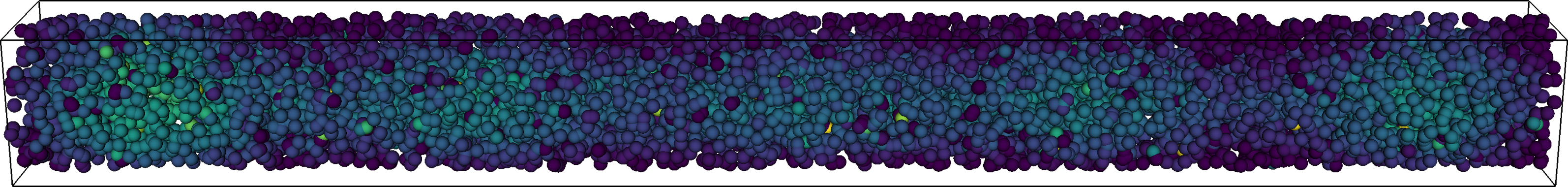}} \\
  \subfloat{\includegraphics[width=\textwidth]{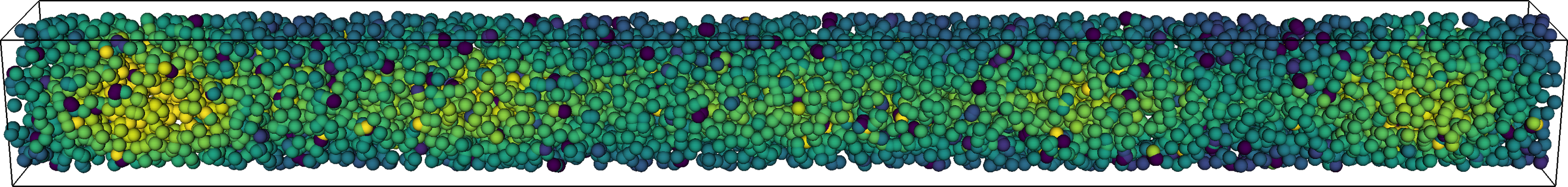}} \\
  \subfloat{\includegraphics[width=\textwidth]{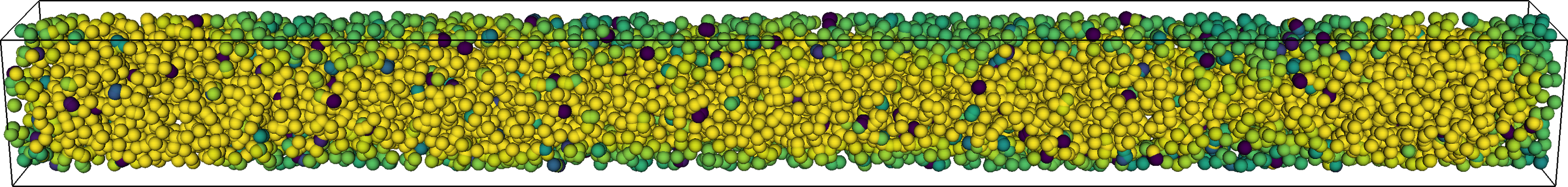}} \\
  \caption{\label{fig:tubes}
    Coloration of $\abs{\tilde{\rho}_{01}}$ at $t_1 = \SI{-0.05}{\pico\second}$, $t_2 = \SI{0}{\pico\second}$, $t_3 = \SI{0.05}{\pico\second}$, and $t_4 = \SI{0.10}{\pico\second}$ relative to the peak of a \SI{1}{\pico\second}-wide pulse.
    \num{10000} \qds{} randomly distributed throughout a $\SI{0.2}{\micro\meter} \text{ (radius)} \times \SI{4}{\micro\meter}$ cylinder oriented along $\vb{k}_L$ demonstrate near-field the effects of \cref{fig:density stats} as distinct, outlying bright/dark \qds{}.
    Additionally, the size of the system allows for wavelength-scale phenomena that appear here as five standing regions of enhanced polarization.
    (Note that we model \qds{} as point objects; the size of the spheres here has no physical interpretation.)
  }
\end{figure*}

\begin{figure}
  \centering
  \begin{tikzpicture}
    \begin{axis}[
        xtick scale label code/.code={$\cdot \lambda$},
        enlargelimits=false,
        axis equal image,
        colormap/viridis,
        colorbar horizontal,
        point meta min=0.46,
        point meta max=0.47,
        colorbar style={
          xtick={0.461, 0.463, 0.465, 0.467, 0.469},
          xticklabels={$\le 0.461$, 0.463, 0.465, 0.467, $0.469 \le$},
          at={(0,1.5)},
          anchor=north west,
        }
      ]
      \addplot graphics [xmin=0,xmax=3,ymin=0,ymax=1] {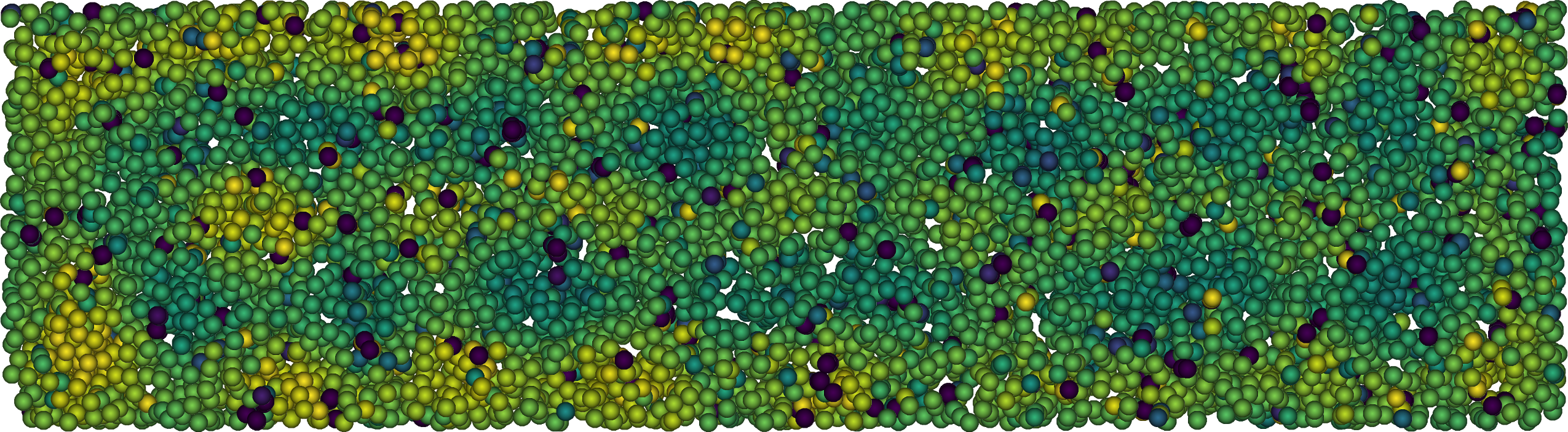};
    \end{axis}
  \end{tikzpicture}
  \caption{\label{fig:wide plate}
    Coloration of $\abs{\tilde{\rho}_{01}}$ for \num{10000} \qds{} arranged in a finite planar geometry (in units of $\lambda$).
    The slab displays a prominent polarization pattern \SI{1.25}{\pico\second} after the peak of a \SI{1}{\pico\second} pulse.
  }
\end{figure}

\Cref{fig:tubes} depicts the evolution of $\abs{\rho_{01}(\vb{r})}$ as an indicator of $\tilde{\vb{P}}$ for a cylinder containing \num{10000} \qds{}.
The cylinder has a radius of \SI{0.2}{\micro\meter} and a length of \SI{4}{\micro\meter}, and the incident $\vb{k}_L$ lies along the cylindrical axis (again perpendicular to $\vb{d}$ so as to maximize the long-distance interaction between \qds{}).
This simulation captures the suppression effects of \cref{fig:density stats,fig:nearfield box} as a small number of \qds{} remain in an unexcited state while the cylinder polarizes around them.
Additionally, due to the length of the cylinder, larger regions of enhanced polarization begin to appear as the system polarizes---an effect we did not observe in sub-wavelength structures.
We liken these nodes to standing waves in a cavity that arise from the far-field interaction term of \cref{eq:radiated envelope}.
As the pulse varies little over the length of the cylinder, identical simulations run without interactions (i.e.~$\mathcal{Z} = 0$ everywhere) produce homogeneous polarization distributions.
Reducing the simulation to a planar geometry (\cref{fig:wide plate}) preserves both the short-range (dark, adjacent \qds{}) and long-range (regions of enhanced/diminished polarization) phenomena observed in \cref{fig:tubes}.

\subsection{Inhomogeneous broadening}

\begin{figure*}
  \centering
  \usepgfplotslibrary{colorbrewer}
\usepgfplotslibrary{groupplots}

\begin{tikzpicture}[]
  \begin{groupplot}[
      group style={
          group name=my plots,
          group size=2 by 2,
          xlabels at=edge bottom,
          xticklabels at=edge bottom,
          ylabels at=edge left,
          yticklabels at=edge left,
          vertical sep=2pt,
          horizontal sep=2pt,
      },
      width=0.5\textwidth,
      height=0.618034\columnwidth,
      %xlabel={Length (\si{\micro\meter})},
      %xmin = 0, xmax = 30,
      %ymin = 0, ymax = 1,
      %minor x tick num={4},
      %minor y tick num={4},
      enlargelimits = 0.03,
      legend pos=south east,
      legend style={empty legend},
      cycle list/Set1-3
  ]
    \nextgroupplot[]
      \addplot[] graphics[xmin=-2.2, xmax=2.2, ymin=0.32, ymax=0.37] {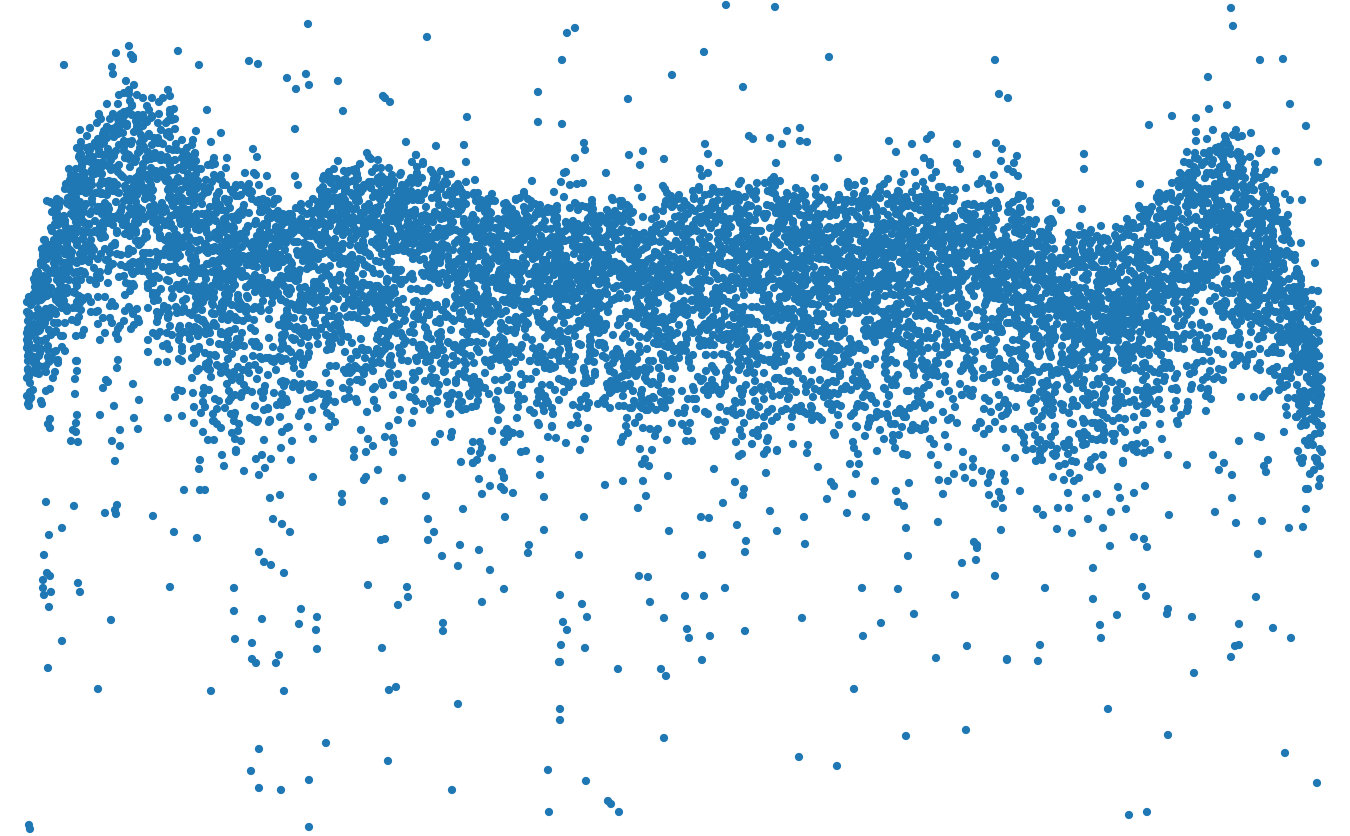};
      \addlegendentry{\SI{0}{\milli\eV}}

    \nextgroupplot[]
      \addplot[] graphics[xmin=-2.2, xmax=2.2, ymin=0.32, ymax=0.37] {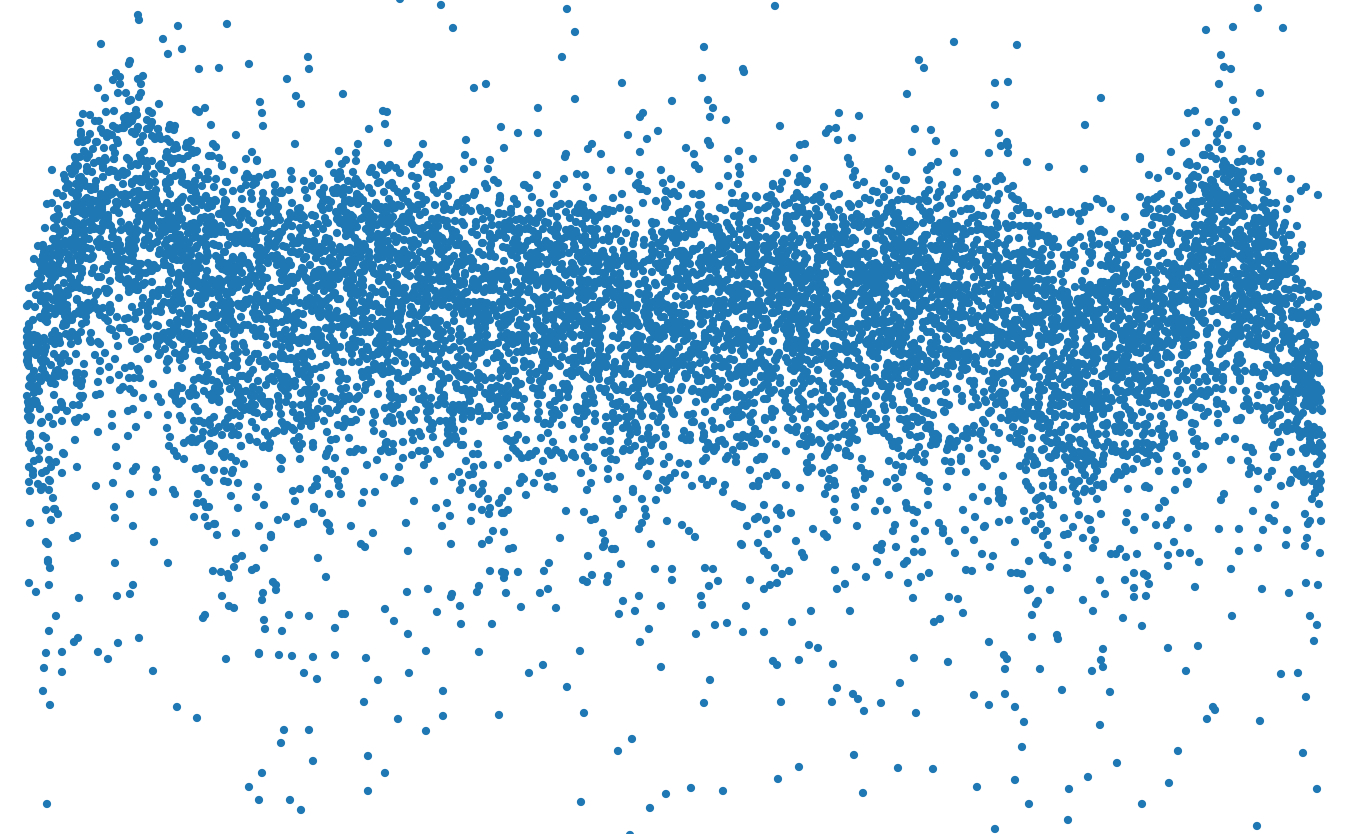};
      \addlegendentry{\SI{0.2}{\milli\eV}}

    \nextgroupplot[]
      \addplot[] graphics[xmin=-2.2, xmax=2.2, ymin=0.32, ymax=0.37] {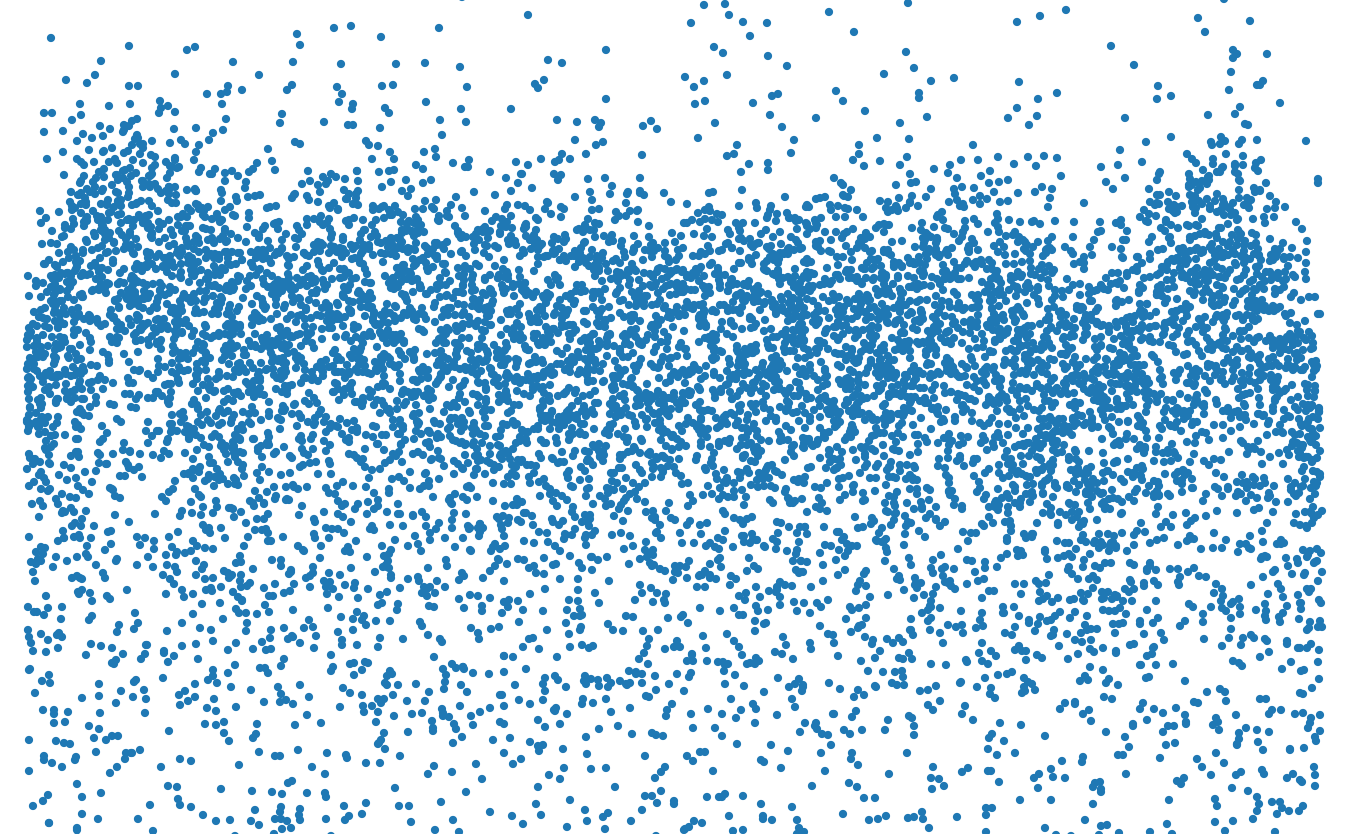};
      \addlegendentry{\SI{0.4}{\milli\eV}}

    \nextgroupplot[]
      \addplot[] graphics[xmin=-2.2, xmax=2.2, ymin=0.32, ymax=0.37] {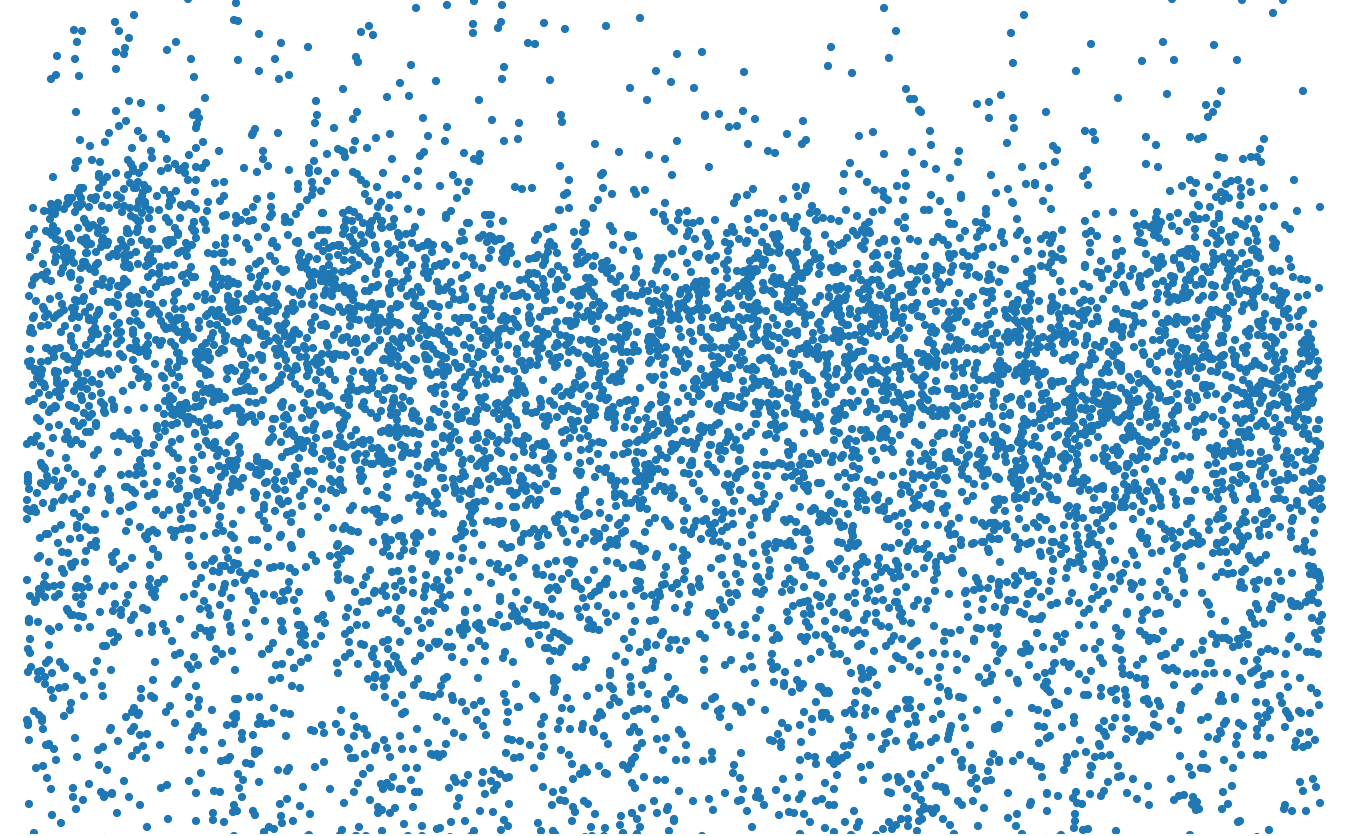};
      \addlegendentry{\SI{0.6}{\milli\eV}}

  \end{groupplot}

\end{tikzpicture}
  \caption{\label{fig:broadened} 
    $z$-distribution of polarization $\abs{\tilde{\rho}_{01}}$ for the geometry in \cref{fig:tubes}.
    In each simulation, \num{10000} \qds{} had random Gaussian noise (with width parameters of \SIlist{0.0;0.2;0.4;0.6}{\milli\eV}) added to a resonant $\hbar \omega_0 = \SI{1500}{\milli\eV}$.
    The induced long-range patterns in the polarization remain for mild detunings $\lesssim \SI{0.5}{\milli\eV}$ but have completely washed out at \SI{0.6}{\milli\eV}.
    Additionally, each simulation displayed the characteristic near-field coupling effects of \cref{fig:density stats} (not visible here).
  }
\end{figure*}

Thusfar we have investigated only homogeneous systems (i.e. an identical $\omega_0$ for every \qd{} in the system).
To probe the effects of interaction-independent inhomogeneities (possibly occuring due to some experimental variation in \qd{} sizes), \cref{fig:broadened} presents four simulations with normally-distributed transition frequencies characterized by a width parameter.
In simulations with mild detuning, we observe a distribution pattern characteristic of the enhancement phenomenon seen in \cref{fig:tubes}.
More variation in the detuning distribution, however, quickly serves to destroy these phenomena leaving only ``pulse-driven'' effects.

\section{\label{section:conclusion}Conclusions \& future work}
Here we developed a robust, fine-grained algorithm to solve for the dynamics of an ensemble of \qds{} that couple in response to external light fields.
By making use of an integral equation kernel to propagate radiated fields, our model facilitates simulations of thousands of \qds{} in three dimensions with accurate bookkeeping of both near and far radiation fields.
Our simulations predict a suppression effect between adjacent \qds{} that screens out the incident laser pulse and we interpret this effect as a dynamical detuning that shifts the effective $\omega_0$ of the affected \qds{}.
Moreover, we observe additional oscillatory behavior and localization effects in larger clusters of particles.
These effects could prove useful to identify optically quantum dot ``molecules'' in an extended sample by detecting residual localized polarization following integral $\pi$ pulse(s)---we expect that an experimental $\pi$-pulse calibrated to a single \qd{} with a scanning-type polarization measurement~\cite{Asakura2013} would reveal signatures of these effects in dense samples.
Finally, in larger systems of densely-packed \qds{}, we see significant localization that present as regions of enhanced polarization over length scales comparable to that of the incident wavelength.
These effects persist in simulations with other extended geometries and in simulations with inhomogeneously-broadened transition frequencies, though the effects quickly disappear with only a few $\si{\milli\eV}$ detuning.

Semi-classical approaches can describe some superradiant effects within a continuum formulation~\cite{Gross1982,PhysRevA.4.302,PhysRevA.4.854}.
First predicted in 2005~\cite{Temnov2005} and observed in 2007~\cite{Scheibner2007}, superradiant effects in \qd{} ensembles have since spurred theoretical analyses into cooperative radiation mechanisms~\cite{Temnov2009,Chen2008}.
While our semiclassical approach accounts for collective effects due to the secondary field emission from \qds{}, we do so in the Hamiltonian term on the right hand side of \cref{eq:liouville} and not in the $\hat{\mathcal{D}}\qty[\hat{\rho}]$ dissipator.
In future work, we plan to extend our microscopic approach to include collective dissipation effects so as to better model superradiant phenomena.
We expect that our approach---when extended to systems containing a larger number of \qds{}---will aid in investigating the role of many-dot interactions in systems such as nanolasers~\cite{jahnke2016giant} that exploit these phenomena.
Unfortunately, the na\"ive $\mathcal{O}\qty(N_s^2)$ interaction calculation presented here hampers attempts to extend these calculations to systems with $N_s \gg 10^4$.
Our ongoing research includes the development of accelerated computational techniques that exploit the structure of $\mathcal{Z}$ to reduce the big-$\mathcal{O}$ complexity of \cref{eq:zmatrix}.
Additionally, the technique presented here readily extends to model atomic, molecular, and semiconductor systems with richer structure (e.g.\ systems with energy degeneracies or biexcitonic transitions).

\acknowledgments
We gratefully acknowledge support from the National Science Foundation grant ECCS-1408115, and extend our thanks to the developers of Eigen~\cite{Eigen}, VisIt~\cite{VisIt}, and NumPy/SciPy~\cite{NumPy,SciPy} for the software used in our simulation and analysis.
\vspace{.5 cm}

\bibliography{random_lasing_reflist}

\end{document}